\documentclass[letterpaper]{article} 
\usepackage[preprint]{aaai2027}  
\usepackage[hyphens]{url}  
\usepackage{graphicx} 
\usepackage{natbib}  
\usepackage{caption} 
\usepackage{amsmath}
\usepackage{amssymb}
\usepackage{array}
\usepackage{tabularx}
\usepackage{multirow}
\usepackage{booktabs}
\usepackage{tikz}
\usepackage{placeins}
\usetikzlibrary{arrows.meta,calc,positioning}

\newcommand{\valid}{\mathrm{Valid}}
\newcommand{\correct}{\mathrm{Correct}}
\newcommand{\answer}{\mathrm{Answer}}
\newcommand{\poll}{\mathrm{poll}}
\newcommand{\speak}{\mathrm{speak}}
\newcommand{\ind}{\mathbb{1}}

\title{What Confidence Routing Is Actually Doing: Auditing Routing, Calibration, and Commitment in Multi-Agent Deliberation}

\author{
    Jingyan Jiang,
    Huihuo Zheng,
    Rajeev Thakur,
    Chih-Hsuan Yang\textsuperscript{*}
}
\affiliations{
    Argonne National Laboratory, Lemont, IL 60439, USA\\
    \{jingyan.jiang, huihuo.zheng, thakur, bellayang\}@anl.gov\\
    \textsuperscript{*}Corresponding author: bellayang@anl.gov
}

\begin{document}

\maketitle

\begingroup
\renewcommand{\thefootnote}{}%
\footnotetext{Preprint. This version has not undergone peer review.}%
\endgroup
\setcounter{footnote}{0}

\begin{abstract}
A common multi-agent design asks agents to report confidence and lets the highest-scoring agent speak next, implicitly using one scalar both to route the conversation and to estimate uncertainty. We audit this confidence-routed broadcast protocol by separating three trace-level questions: whether it selects the right candidate (\emph{routing}), whether reported confidence behaves like a probability (\emph{calibration}), and whether the selected agent publicly states the answer that won the turn (\emph{commitment}). Our primary study covers 4{,}181 \texttt{gpt-oss-120b} olympiad-math traces; we repeat the audit on a $2\times2$ actor$\times$benchmark grid that adds \texttt{gemma-4-31B-it} and a biology multiple-choice benchmark. In the primary cell, confidence discriminates correct from wrong candidates (AUROC 0.72) but is strongly overconfident (79\% mean stated confidence versus 52\% accuracy). A cross-fitted, tier-stratified isotonic procedure reduces Expected Calibration Error from $0.278$ to $0.008$ on held-out candidates, but it does not recover missing discrimination: raw AUROC is only $0.537$ and $0.440$ in the two Gemma cells. Routing is likewise setting-dependent. Fixed routers differ by at most 1.1~percentage points on \texttt{gpt-oss}/math, whereas raw-confidence argmax performs 5.6 and 11.2~points below random-valid selection in the Gemma cells. Commitment is distinct again: in the primary cell, poll and spoken answers diverge in 20.4\% of valid pairs, 62.4\% of those revisions are fresh generations, and the unconditional correctness shift is $-1.7$~points; the other three cells instead range from $+0.9$ to $+12.2$~points. The transferable lesson is procedural: routing discrimination, probability calibration, and public commitment must be measured separately before raw confidence is used for deployment decisions.
\end{abstract}

\section{Introduction}

Multi-agent deliberation replaces a single LLM call with a protocol \citep{du2023improving,wang2024mixtureofagents,khan2024debating}. Agents propose answers, see each other, and update over rounds. A common design asks each agent for a verbalized confidence score and lets the most confident agent speak next. Verbalized confidence is easy to elicit from black-box LLMs and can carry useful signal about correctness \citep{lin2022teaching,tian2023just,xiong2024uncertainty}.
That design is appealing because it seems to turn one cheap self-report into both a speaker-selection rule and an uncertainty estimate.

This design also creates a measurement gap. The confidence score and candidate answer come from one prompt and format. The public message comes from a different one. In our broadcast traces, each agent first answers a structured JSON poll with a candidate and a confidence score, the router picks the highest-confidence agent, and only then does that agent write a free-form public message. The answer used to allocate the turn is therefore not always the answer that enters the transcript. Format constraints can change generated outputs \citep{tam2024speakfreely,sclar2024sensitivity}, and verbalized confidence depends on field order and elicitation context \citep{tian2023just,seo2025advice,jang2025verbalized}.

Recent audits show that multi-agent debate does not reliably outperform strong single-agent baselines under careful accounting \citep{zhang2025overvaluing,smit2023mad,kaushal2026deliberationbench,tran2026singleagent,wu2025reallydebate}, that coordination errors deserve separate analysis \citep{cemri2025mast}, and that self-correction without external feedback can fail \citep{huang2024selfcorrect,kamoi2024selfcorrection}. These results motivate finer-grained evaluation: final accuracy alone cannot distinguish missing candidate content, an unhelpful routing decision, a miscalibrated score, or a change between the private poll and the public answer. We make three contributions. First, we define routing, calibration, and commitment as separately measurable trace objects with explicit denominators. Second, we show why discrimination and probability scale must not be conflated: in the primary \texttt{gpt-oss}/math cell, a cross-fitted isotonic procedure reduces Expected Calibration Error from 0.278 to 0.008 on held-out candidates, while the cross-model grid shows that raw discrimination and routing value are setting-dependent. Third, we provide a compact trace-audit card so that future evaluations can report these axes side by side (Appendix~\ref{app:trace_audit_card}).

Our primary audit examines broadcast math-reasoning traces from \texttt{openai/gpt-oss-120b} under a runtime confidence-argmax router; \S\ref{sec:ablation} repeats the audit with \texttt{google/gemma-4-31B-it} and a biology benchmark. The primary traces come from a broader verifier-gated evaluation that also ran non-collaborative comparators on the same 4{,}181 problems. Broadcast has a higher same-task pass rate but a large cost premium (12.8$\times$ tokens, Appendix~\ref{app:single_agent_context}), which makes a narrower question important: \emph{what is the confidence score actually buying?} We answer that question by separating candidate availability, selection quality, probability scale, and the stability of the selected answer at the public speak step.

The audit answers that question in five steps.
\begin{itemize}
  \item \textbf{We separate three things usually treated as one.} A confidence-routed turn has a routing decision, a confidence score, and a public answer. We define separate metrics for each, with explicit denominators.
  \item \textbf{Routing differences are small.} On gpt-oss/math, five fixed routers fall within 1.1~pp of one another and 4.8--5.9~pp below an oracle that selects a correct candidate whenever one exists.
  \item \textbf{The score ranks better than it calibrates, and the speaker does not always keep the answer that won the turn.} Confidence separates correct from wrong above chance (AUROC 0.721) but is badly overconfident (0.790 stated vs.\ 0.518 accuracy). The selected agent revises its polled answer in 20.4\% of valid pairs, 62\% of them novel regenerations, and on gpt-oss/math the net shift is $-1.7$~pp, though its sign is setting-dependent across the grid.
  \item \textbf{Candidate availability and routing quality are distinct.} A same-trace outcome regression associates success most strongly with how often a block contains a correct candidate. We use this as descriptive localization, while the router counterfactuals quantify selection headroom directly.
  \item \textbf{The decomposition transfers, but individual effects vary.} On a $2\times2$ grid adding a \texttt{gemma-4-31B-it} actor and a biology benchmark (\S\ref{sec:ablation}), scale repair recurs, raw discrimination varies, and confidence argmax can be worse than random-valid selection. Raw confidence should therefore be audited before it is used for routing.
\end{itemize}

\section{Protocol under audit}\label{sec:protocol}

\paragraph{Broadcast deliberation.}
Each trace is a broadcast deliberation over one math problem, in which all agents see the same public transcript, unlike peer-to-peer or star topologies. We focus on broadcast because it is the simplest topology in which the routing decision, the confidence score, and the public commitment are all observable on the same trace. At each round, every active agent reads the transcript and returns a structured poll with a confidence score and a candidate answer. The system picks one agent by runtime confidence argmax with a deterministic tie rule. The picked agent then writes a free-form public message, whose answer may match the polled one, differ from it, or be less extractable. Between discussion rounds, an approval sub-protocol can endorse a current system answer, which later appears in the trace as the system incumbent. Appendix~\ref{app:protocol_details} summarizes the exact prompt surfaces, visibility model, and configuration choices that create this measurement gap.

\begin{figure*}[t]
\centering
\begin{tikzpicture}[
  font=\small,
  stage/.style={
    draw=black,
    rounded corners=3pt,
    text width=3.1cm,
    minimum height=2.25cm,
    align=center,
    inner sep=6pt
  },
  transcriptstage/.style={stage, fill=teal!8},
  pollstage/.style={stage, fill=blue!8},
  routerstage/.style={stage, fill=orange!10},
  speakstage/.style={stage, fill=violet!8},
  flow/.style={
    -{Latex[length=2.5mm]},
    thick
  },
  auditnote/.style={
    draw=black,
    rounded corners=3pt,
    fill=white,
    text width=8cm,
    align=center,
    inner sep=5pt,
    font=\small
  }
]
\node[transcriptstage] (transcript) at (0,0) {\textbf{Public transcript}\\[2pt]$h_t$ visible to all agents};
\node[pollstage] (poll) at (3.75,0) {\textbf{Structured poll}\\[2pt]each agent returns\\$(c_{i,t}, s_{i,t})$};
\node[routerstage] (router) at (7.5,0) {\textbf{Confidence router}\\[2pt]selects speaker\\$i^*_t=\arg\max_i s_{i,t}$};
\node[speakstage] (speak) at (11.25,0) {\textbf{Free-form speak}\\[2pt]selected agent emits $u_{i^*,t}$\\with extracted $\hat c_{i^*,t}$};

\draw[flow] (transcript.east) -- (poll.west);
\draw[flow] (poll.east) -- (router.west);
\draw[flow] (router.east) -- (speak.west);

\node[auditnote] (gap) at ($(poll.south)!0.5!(speak.south) + (0,-1.35)$)
{\textbf{Audit target:} the protocol uses the structured poll candidate $c_{i,t}$ to allocate the turn, but the free-form spoken candidate $\hat c_{i^*,t}$ is what enters the public transcript.};

\draw[densely dashed, thick] ($(poll.south)+(0,-0.05)$) -- ($(poll.south)+(0,-0.5)$);
\draw[densely dashed, thick] ($(speak.south)+(0,-0.05)$) -- ($(speak.south)+(0,-0.5)$);
\draw[densely dashed, thick] ($(poll.south)+(0,-0.5)$) -- ($(speak.south)+(0,-0.5)$);
\end{tikzpicture}
\caption{The audited protocol has two answer-bearing calls. In round $t$, agent $i$ returns a structured poll with candidate answer $c_{i,t}$ and stated confidence $s_{i,t}$. The router selects the highest-confidence agent $i^*$, which then emits a free-form public message $u_{i^*,t}$ with extracted spoken candidate $\hat c_{i^*,t}$. The poll candidate is used only to allocate the turn, whereas the spoken candidate enters the public transcript. The audit treats the routing decision, the calibration of $s_{i,t}$, and the relationship between $c_{i^*,t}$ and $\hat c_{i^*,t}$ as three distinct objects.}
\label{fig:protocol}
\end{figure*}
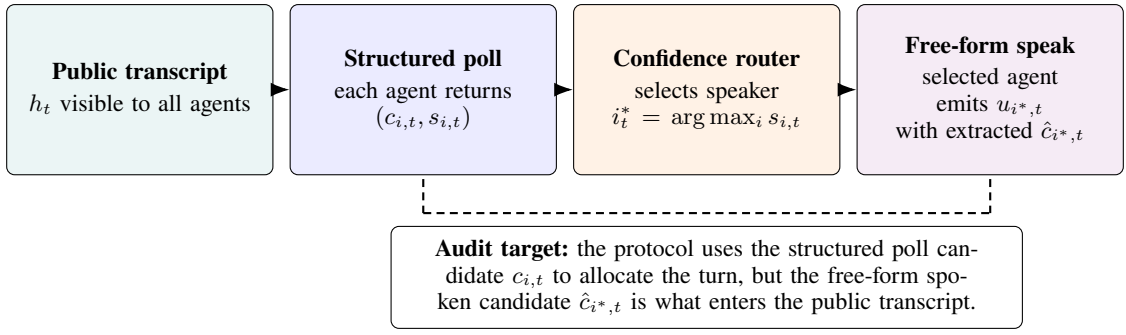

\paragraph{Three audit objects.}
The protocol exposes three measurement targets. \emph{Routing}: does the argmax pick a correct candidate when one exists, and how does it compare to other routers? \emph{Calibration}: is the score a useful rank for correctness, and does its stated value behave like a probability? \emph{Commitment}: does the picked agent's public message preserve the polled answer, and when it does not, does correctness go up or down? A within-broadcast outcome regression then asks which of the three predicts whether the broadcast solves the problem.

\section{Metrics and audit machinery}\label{sec:metrics}

All correctness-based metrics are validity-gated. We mark a poll or speak candidate valid when a concrete answer can be extracted, and define correctness only on valid candidates. Validity gating matters because extraction failures can otherwise make routing or confidence look worse for reasons unrelated to the underlying reasoning. Equivalence uses numeric and symbolic fast paths where possible, with an LLM judge for residual cases.

\paragraph{Routing metrics.}
A poll block is one routing decision point: the set of per-agent poll candidates for one turn, from which the router selects a candidate $c^*$. For routing, we track five plain-language quantities: selected-candidate accuracy, hit rate when a correct candidate exists, missed-correct rate, within-block confidence-versus-correctness alignment, and the confidence tie rate. We retain the abbreviations CRA, HRWP, MCR, PCCA, and CTR for tables. PCCA and AUROC use different stratifications and are not interchangeable. A glossary of all abbreviations is in Appendix~\ref{app:notation}.

\paragraph{Validity-gating accounting.}
Validity gating changes the denominators, so we report the gating rates too. Across all 11{,}122 selected blocks the selected candidate is invalid 14.9\% of the time, but on the subset of blocks that contain a \emph{correct alternative} (a valid correct candidate somewhere in the block) this falls to 0.7\%, so most invalid selections are a block-quality issue rather than a routing mistake.

\paragraph{Calibration metrics.}
We report Expected Calibration Error (ECE), which measures the gap between stated confidence and empirical accuracy across equal-width bins, together with the Brier score, mean stated confidence vs.\ accuracy, and AUROC of confidence against poll-candidate correctness. The Brier score is the mean squared error of the probability forecast, a proper scoring rule that a rank-preserving recalibration cannot game \citep{brier1950verification}. Confidence intervals for these pooled calibration metrics use a cluster bootstrap at the problem-cluster level.

\paragraph{Commitment metrics.}
For selected speak events, we compare the polled candidate $c^\poll$ with the extracted spoken candidate $c^\speak$, asking how often poll and speak differ, how often that difference helps versus hurts, and what the net correctness shift is. We retain the abbreviations PSDR, PSIR, and PSDRg, and use separate notation for conditional and unconditional net shifts. For the headline pair-micro estimates, let $\mathcal{P}$ be the pooled set of valid labeled selected pairs $(c^\poll, c^\speak)$ in an analysis cell, and let $\mathcal{D}\subseteq\mathcal{P}$ be its divergent subset ($\answer(c^\poll)\not\equiv\answer(c^\speak)$). Then
\begin{align*}
\text{PSDR} &= |\mathcal{D}|/|\mathcal{P}|, \quad
\text{NPCS}_{\mid\mathcal{D}} = \text{PSIR} - \text{PSDRg}, \\
\text{PSIR} &= \tfrac{1}{|\mathcal{D}|}\big|\{\mathcal{D} : \neg\correct(c^\poll) \land \correct(c^\speak)\}\big|, \\
\text{PSDRg} &= \tfrac{1}{|\mathcal{D}|}\big|\{\mathcal{D} : \correct(c^\poll) \land \neg\correct(c^\speak)\}\big|, \\
\text{NPCS}_\text{unc} &= \tfrac{1}{|\mathcal{P}|}\!\!\sum_{\mathcal{P}} \big[\correct(c^\speak) - \correct(c^\poll)\big].
\end{align*}
The conditional $\mathrm{NPCS}_{\mid\mathcal D}$ is micro-averaged over divergent pairs; $\mathrm{NPCS}_\text{unc}$ is micro-averaged over all valid pairs. The outcome regression does not use conditional PSIR or PSDRg. It uses per-trace unconditional wrong-to-right and right-to-wrong event rates over selected speak events, so a trace with no such event contributes zero to the corresponding predictor.

\paragraph{LLM-judge audit.}
Residual equivalence cases go to \texttt{gpt-oss-120b} in a self-judge configuration, where the same model family both deliberates and judges. Because self-preference is a concern \citep{wataoka2024selfpreference}, we attempt cross-judge re-adjudication of all 4{,}379 non-trivial residual pairs from tiers 02--09 with \texttt{google/gemma-4-31B-it} and \texttt{google/gemma-4-E4B-it}. We then recompute commitment metrics over the full 9{,}263 valid-pair population in those tiers, using each judge's labels for residual cases. Conditional NPCS is $-0.083$, $-0.061$, and $-0.060$ under the three judge variants. This checks sensitivity to residual-case adjudication; it does not audit fast-path equivalence decisions or errors shared by all judges. Per-judge $\kappa$ and confusion counts, plus metric-behavior sanity checks under within-block permutation and noisy-oracle constructions, are in Appendices~\ref{app:sanity} and~\ref{app:judge_audit}.

\paragraph{Uncertainty and dependence.}
A \emph{problem cluster} groups questions that share a corpus-shard position and serves as a single resampling unit (not a set of repeated attempts at one question), so within-cluster dependence does not inflate our confidence. Pooled CIs use a problem-cluster bootstrap over 421 clusters, and regressions use cluster-robust standard errors at the same level \citep{cameron2015practitioners}. The two grouping keys play distinct roles: \texttt{problem\_uid} is the leakage-control unit for calibration cross-fitting, so a question never appears in both folds, and \texttt{cluster\_id} is the resampling and dependence unit. Full reporting parameters are in Appendix~\ref{app:protocol_details}.

\section{Experimental setting}\label{sec:experimental}

The audit corpus is broad in source but fixed in protocol. It contains broadcast traces from ten math-reasoning difficulty tiers drawn from olympiad-style sources (HMMT, olympiad shortlists, APMO, Putnam-style, and related material in Omni-MATH-style metadata \citep{gao2024omnimath}), in a task format close in spirit to benchmarked math problem solving \citep{hendrycks2021math}. The audited slice spans 64 named source collections and 4{,}181 problems, with \texttt{openai/gpt-oss-120b} deliberators. Those problems yield 4{,}181 traces, 23{,}391 valid labeled poll candidates, 11{,}122 selected poll blocks, 9{,}436 valid labeled selected poll-speak pairs, and 421 problem clusters (mean about 10 traces per cluster). The same corpus supports routing measurements at the block level, calibration at the candidate level, and commitment measurements at the selected-pair level. The broader verifier-gated evaluation that produced these traces also evaluated non-collaborative comparators on the same 4{,}181 problems, so we later report same-task performance and cost context (Appendix~\ref{app:single_agent_context}). Code, data, and corpus summaries are not distributed in this arXiv source package; Section~\ref{app:analysis_artifacts} states the availability boundary.

To turn raw traces into audit labels, we extract boxed answers, the standard math-benchmark marker for a final answer, or otherwise parseable answers from both poll and speak responses. Each candidate's correctness is decided by the runtime verifier against the dataset gold answer. For the per-problem reference label we take a verifier-confirmed final answer where one exists and the dataset gold answer otherwise, so the reference is gold-grounded in either case. No problem-solving traces were rerun for any post-hoc analysis.

\section{Routing differences are small in this setting}\label{sec:routing}

This section reports the primary gpt-oss/math cell. We compare routers over the 9{,}653 all-routeable poll blocks, where every compared router has at least one valid candidate to select from (the 11{,}122 selected blocks minus 1{,}469 with no valid labelable candidate). On these blocks, five fixed routers fit inside a band of about 1 percentage point (49.0--50.1\%, Table~\ref{tab:routing}), where confidence-weighted vote sums stated confidence within each distinct answer and picks the largest total. The oracle ceiling, which succeeds whenever any valid labelable candidate in the block is correct, reaches only 54.9\%, so every router lies 4.8--5.9~pp below the best result achievable given the candidates the pool actually produced. The deterministic runtime tie rule is likewise minor: replacing it with uniform randomization over top-confidence ties moves expected success by only 0.4~pp. \emph{Within this router class and on this pool, the routing rules differ little from one another. The larger constraint is whether the block contains a correct candidate at all.} This pool is homogeneous, three gpt-oss peers at temperature $0$, so within-block diversity is low and a real routing choice exists only in the mixed-correctness blocks (about $1{,}184$ of $11{,}122$). Interchangeability is thus a within-pool statement, not a cross-actor one: the ablation in \S\ref{sec:ablation} shows that on a different actor the confidence scalar can be actively harmful to route on.

Confidence-weighted voting is the aggregation rule at the heart of round-table consensus schemes such as ReConcile \citep{chen2024reconcile}. It is numerically best here, but only by 0.4~pp over the deployed argmax router. Learned routing is an active target in its own right \citep{yue2025masrouter,liu2024dylan}. We run the offline analogue as a stress test. A learned router, a logistic model trained to predict candidate correctness from routing-time features and cross-fitted by \texttt{problem\_uid}, confirms the interchangeability: it selects a correct candidate on 49.9\% of blocks, tied with argmax (95\% CIs $[0.478,0.521]$ vs.\ $[0.477,0.519]$), even though as a candidate-level classifier its pooled AUROC is far higher than the raw score ($0.806$ vs.\ $0.721$). The reason is structural: its strongest feature, the deliberation round index, is constant across candidates within a block and cannot break the within-block choice. A better correctness predictor does not become a better router, because pool content, not the selection rule, caps what any router can achieve.

\begin{table}[t]
\centering
\small
\setlength{\tabcolsep}{4pt}
\begin{tabular}{lcc}
\toprule
Router & \shortstack{Expected\\success} & \shortstack{Gap to\\oracle} \\
\midrule
Runtime confidence argmax & 0.497 & 0.052 \\
Top-confidence tie-random & 0.493 & 0.055 \\
Random valid candidate & 0.490 & 0.059 \\
Exact-answer majority vote & 0.496 & 0.052 \\
Confidence-weighted cluster vote & 0.501 & 0.048 \\
Learned router ($P(\text{correct})$) & 0.499 & 0.050 \\
Oracle over valid candidates & 0.549 & 0.000 \\
\bottomrule
\end{tabular}
\caption{Router counterfactuals on the gpt-oss/math cell, over 9{,}653 all-routeable poll blocks. Five fixed-rule routers cluster within 1.1~pp, and a learned router trained to predict candidate correctness lands in the same band. The oracle ceiling is 4.8--5.9~pp above them all. For this broadcast protocol, the routing rule matters less than whether the block contains a correct candidate.}
\label{tab:routing}
\end{table}

The denominator accounting strengthens the same point. In 13.2\% of blocks no agent produces a valid candidate at all, and the selected-invalid rate (14.9\% overall) falls to 0.7\% once a correct alternative exists, so most invalid selections trace to blocks where the pool supplies no usable option. Within the 1{,}184 mixed-correctness blocks, PCCA is 0.570, above its 0.500 no-signal reference; the confidence-tie rate is 0.327 over 2{,}079 correct--wrong comparisons (Appendix~\ref{app:body_floats}). The router captures a correct candidate in 90.6\% of the 5{,}295 possible-hit blocks; the separate missed-correct rate is 5.8\% over 7{,}970 eligible blocks. These quantities use different denominators.

Even if routing is good enough when a correct candidate exists, does the confidence value itself mean what its numeric scale claims? The next section separates two uses of the score, ordering candidates and estimating the probability that a candidate is correct.

\section{Confidence is a poor probability, but the gap is cheaply fixable}\label{sec:calibration}

Confidence separates stronger candidates from weaker ones better than it estimates their absolute probability of being correct. This ordering ability is the \emph{rank signal}: whether higher confidence tends to score correct candidates above wrong ones, measured by AUROC, where 0.5 is no signal. Over 23{,}391 valid candidates, AUROC for confidence against poll-candidate correctness is 0.721 [0.708, 0.733], but mean stated confidence is 0.790 against accuracy of 0.518: a 27.2~pp overconfidence gap [$+25.2$, $+29.1$], with Expected Calibration Error 0.278 [0.270, 0.286] and Brier score 0.302 [0.290, 0.313] (raw row of Table~\ref{tab:recalibration}). Higher confidence usually does point toward better candidates, but the score's numeric scale overstates how often those candidates are actually correct.

This split between rank and scale becomes even clearer across tiers. Per-tier ECE rises sharply from $0.10$ (tier 01) into the $0.31$--$0.41$ range on tiers 06--10, while the rank signal weakens more gradually, and the pooled confidence-correctness regression has a strongly negative score-by-difficulty interaction, so confidence becomes flatter and less decision-useful on harder problems (Appendices~\ref{app:pertier} and~\ref{app:regression}). Whether that reflects intrinsic difficulty or pre-training coverage is a separate question (\S\ref{sec:limitations}) that does not affect the pooled calibration claim.

\paragraph{The probability scale is post-hoc repairable on held-out candidates.}
We fit two standard calibrators: temperature scaling divides confidence logits by a learned scalar \citep{guo2017calibration}, and isotonic regression fits a nondecreasing step function from stated confidence to empirical accuracy. Both use $K{=}2$ cross-fitting by \texttt{problem\_uid}, with fold assignment stratified by tier, so each training fold produces its own calibrator and no question appears in both fit and evaluation data. Each isotonic map is constrained to be nondecreasing: it cannot reverse a strict score ordering among values transformed by that map, but it can merge score levels and create ties. On held-out candidates, isotonic reduces ECE from 0.278 to 0.008 and the Brier score from 0.302 to 0.208; candidate-level AUROC changes only from 0.721 to 0.720 (Table~\ref{tab:recalibration}). We foreground Brier because it is a proper scoring rule, whereas equal-width ECE is biased and post-hoc isotonic ECE is partly mechanical. Raw ECE is stable across binning schemes (0.272--0.279, equal-width and equal-mass at 10/15/20 bins), isotonic ECE remains lower under equal-mass bins (0.028--0.047), and the Brier drop is binning-independent. Temperature scaling, with one parameter, reduces ECE to 0.142. These results show that a flexible monotone procedure improves the probability scale on held-out data; they do not show that calibrated rerouting would preserve speaker choices or task accuracy.

\begin{table}[t]
\centering
\small
\setlength{\tabcolsep}{3pt}
\begin{tabular}{lcccc}
\toprule
Calibrator & ECE $\downarrow$ & Brier $\downarrow$ & AUROC & mean conf. \\
\midrule
Raw (as reported) & 0.278 & 0.302 & 0.721 & 0.790 \\
Temperature ($T{=}4.35$) & 0.142 & 0.242 & 0.722 & 0.580 \\
Isotonic & \textbf{0.008} & \textbf{0.208} & 0.720 & 0.518 \\
\bottomrule
\end{tabular}
\caption{Post-hoc recalibration on the \texttt{gpt-oss}/math cell, cross-fitted by \texttt{problem\_uid} ($K{=}2$, fold assignment stratified by tier) over 23{,}391 valid poll candidates. ECE uses 10 equal-width bins. The isotonic procedure reduces ECE from 0.278 to 0.008 and Brier score from 0.302 to 0.208; candidate-level AUROC changes only from 0.721 to 0.720. These are held-out candidate-level estimates, not a rerun of the router on calibrated values.}
\label{tab:recalibration}
\end{table}

The held-out calibration can support thresholded analyses such as early stopping, abstention, or cascade gating \citep{kadavath2022know}. In a risk--coverage sweep on the same 23{,}391 candidates, treating the raw score literally overpromises: a threshold of $0.7$ retains 75\% of candidates at 59\% accuracy. After isotonic calibration, the ${\ge}0.7$ subset covers 14\% at 91\% accuracy. This descriptive sweep shows how the probability interpretation changes; it does not establish the effect of calibrated thresholds on an end-to-end policy. Likewise, the outcome regression in \S\ref{sec:linkage} does not show a stable marginal association for selected confidence across sensitivity specifications.

\section{The poll is not a public commitment, and many revisions are fresh regenerations}\label{sec:commitment}

A router can choose the answer that wins the turn and still lose it one step later. That the poll answer and the public answer can differ at all is the paper's most distinctive \emph{measurement} object. Whether that difference helps or hurts is a separate, setting-dependent question. Among 9{,}436 valid labeled selected poll-speak pairs, 20.4\% diverge under the headline \texttt{gpt-oss-120b} equivalence judge ($26$--$27\%$ under the two cross-family Gemma variants, and $20$--$27\%$ across all three under conservative NaN-as-divergent treatment; Appendix~\ref{app:judge_audit}). Divergence is substantial in every grid cell (PSDR 0.11--0.27, Table~\ref{tab:cross_benchmark}), but the net correctness effect has no fixed sign. On \texttt{gpt-oss}/math, repairs occur in 17.4\% of divergent pairs and degradations in 25.7\%, giving $\mathrm{NPCS}_{\mid\mathcal D}=-8.4$~pp [$-11.7$,$-5.0$]. Over all valid labeled pairs, $\mathrm{NPCS}_{\mathrm{unc}}=-1.7$~pp [$-2.4$,$-1.0$]. The other three grid cells instead have positive unconditional shifts from $+0.9$ to $+12.2$~pp, so revision harm is specific to the primary cell. The protocol creates room for this gap because the poll is structured and answer-only, whereas the speak step reopens free-form generation.

Divergence is not automatically harmful. Revision can improve outputs under tool feedback or external scaffolding \citep{shinn2023reflexion,yao2023tree,snell2024scaling}; prior studies also report settings in which intrinsic self-correction without external feedback degrades reasoning \citep{huang2024selfcorrect,kamoi2024selfcorrection}. In our primary cell, a regression in Appendix~\ref{app:regression} associates correct polled candidates with less divergence, with a weaker association on harder problems.

\paragraph{Where the divergent spoken answer comes from.}
We classify each divergent valid spoken answer against what was visible at speak time: the public transcript of prior-round speak messages, and the system-tracked incumbent (the candidate the approval rounds have already endorsed and carried across rounds). Other agents' polls are excluded, since they are not in the speaker's prompt. An answer is \emph{self-revert} if it matches the speaker's own earlier public answer, \emph{peer-echo} if it matches a peer's, \emph{incumbent-pull} if it matches the endorsed system answer, and \emph{novel} if it matches none. When more than one flag fires we assign a single primary class by precedence (self-revert $\succ$ incumbent-pull $\succ$ peer-echo $\succ$ novel). Table~\ref{tab:derivation} reports the decomposition.
\begin{table}[t]
\centering
\small
\setlength{\tabcolsep}{4pt}
\begin{tabular}{lrrr}
\toprule
Primary class & N & \shortstack{\% of\\diverg.} & \shortstack{$\Delta$correct (pp,\\speak $-$ poll)} \\
\midrule
self-revert    & 367 & 19.0 & $-11.4$ \\
incumbent-pull & 120 & \phantom{0}6.2 & $+1.7$ \\
peer-echo      & 238 & 12.4 & $-5.0$ \\
novel          & 1{,}202 & 62.4 & $-9.1$ \\
\bottomrule
\end{tabular}
\caption{Sources of divergent spoken answers ($N=1{,}927$ divergent valid pairs), by precedence-resolved primary class (self-revert $\succ$ incumbent-pull $\succ$ peer-echo $\succ$ novel). Negative $\Delta$correct means the spoken revision is net-harmful on that subset. Overlapping (non-exclusive) match flags are in Appendix~\ref{app:regression}.}
\label{tab:derivation}
\end{table}

Under this protocol's visibility model, the largest primary class is \emph{novel generation}: 62.4\% of divergent spoken answers match no candidate visible in the prior context. Novel revisions have a $-9.1$~pp per-pair shift; self-reverts are more negative per pair at $-11.4$~pp, but the much larger novel class contributes most of the aggregate harm (Table~\ref{tab:derivation}). Incumbent-pull is the only class with a non-negative point estimate ($+1.7$~pp). Across the three residual-case judge variants, conditional NPCS over tiers 02--09 ranges from $-8.3$ to $-6.0$~pp, so its sign is stable under this sensitivity check.

\textbf{Putting this together with the calibration result.} The adoption logit in Appendix~\ref{app:regression} is observational. A visible-majority indicator is associated with adoption, and candidates associated post hoc with higher source-poll confidence are more likely to match the spoken answer after controlling for the listed covariates. Because private poll confidence is not visible to the speaker, this coefficient is not an effect of confidence visibility. Under precedence-resolved labels, 12.4\% of divergences are primary peer echoes; under non-exclusive flags, up to 28.5\% match a prior peer answer. These matches do not identify the fraction causally attributable to peer exposure.

\section{Outcome linkage: what best predicts broadcast success?}\label{sec:linkage}

Not every local audit signal is equally associated with end-to-end broadcast success. We fit an exploratory outcome logit for the log-odds that a trace solves its problem, using z-scored per-trace predictors over 4{,}181 completed traces ($n_{\text{clusters}} = 421$) with cluster-robust standard errors (Table~\ref{tab:linkage}). The lead predictor is the \emph{pool-solvable rate}: the fraction of poll blocks in an evolving trace that contain at least one valid correct polled candidate. It is router-independent within a block, but it is computed from the same completed trace as the outcome, can reflect later feedback and recovery, and is not a pre-treatment variable. It is also not a strict ceiling on end-to-end success, because the speak stage can regenerate a correct answer from an incorrect poll. Other predictors are selected-invalid rate, mean selected confidence, divergence rate (PSDR), unconditional wrong-to-right and right-to-wrong event rates over selected speak events, number of speak events, and tier difficulty. The related selected-correct/routing-slack specification is in Appendix~\ref{app:linkage_decomp}.

\begin{table}[t]
\centering
\small
\setlength{\tabcolsep}{3pt}
\begin{tabular}{lrrr}
\toprule
Predictor (z-scored) & $\beta$ & \shortstack{OR\\per SD} & $p$ \\
\midrule
Pool-solvable rate (router-indep.) & $+4.80$ & $\approx 121$ & $1.3 \cdot 10^{-12}$ \\
Uncond. improvement-event rate & $+1.30$ & $3.68$ & $0.003$ \\
Tier difficulty & $+0.39$ & $1.48$ & $5.1 \cdot 10^{-4}$ \\
Selected-invalid rate & $+0.58$ & $1.78$ & $1.3 \cdot 10^{-5}$ \\
Divergence rate (PSDR) & $+0.15$ & $1.16$ & $0.22$ \\
Uncond. degradation-event rate & $+0.09$ & $1.09$ & $0.91$ \\
Mean selected score & $-0.17$ & $0.84$ & $0.35$ \\
Speak events ($\log 1+\cdot$) & $-1.63$ & $0.20$ & $1.6 \cdot 10^{-21}$ \\
\bottomrule
\end{tabular}
\caption{Exploratory within-broadcast outcome linkage. Logistic regression with cluster-robust SEs by problem cluster ($n=4{,}181$, $n_{\text{clusters}}=421$). Pool-solvable rate has the largest same-trace association. For mean selected confidence, the reported MLE is small and non-significant (OR 0.84, $p=0.35$), while sensitivity refits span OR 0.84--1.11 (Appendix~\ref{app:linkage_decomp}); its sign is therefore not stable.}
\label{tab:linkage}
\end{table}

Pool-solvable rate has the largest coefficient (OR $\approx 121$, 95\% CI $[32,456]$) and nearly separates successful from unsuccessful traces. We interpret this only as descriptive localization: successful completed traces tend to contain correct polled candidates more often. The odds ratio is not a causal effect of increasing pool coverage and does not establish pool content as the unique binding constraint. Router counterfactuals, rather than this regression, quantify the consequences of changing selection rules.

The unconditional improvement-event rate is positively associated with outcome in the reported model (OR $\approx 3.7$), while divergence and degradation-event rates are imprecise. Mean selected confidence has OR 0.84 and $p=0.35$ in the reported cluster-robust MLE, but sensitivity specifications span OR 0.84--1.11, so its sign is not stable. Firth and dropped-pool-solvable refits leave the improvement-event rate positive (OR $3.5$--$9.1$) and give a similar pool-solvable point estimate where included (Appendix~\ref{app:linkage_decomp}). Because the predictors are same-trace summaries and the analyses were not pre-registered, these coefficients are descriptive and hypothesis-generating. The direct deployment evidence comes from the router counterfactuals, held-out calibration evaluation, and measured local poll-to-speak shifts.

\section{Ablation: cross-actor and cross-benchmark}\label{sec:ablation}

The sections above audit one cell: \texttt{openai/gpt-oss-120b} on olympiad math. To test whether the decomposition and its conclusions are specific to that actor or corpus, we re-run the audit on a $2\times2$ grid adding \texttt{google/gemma-4-31B-it} and the LAB-Bench biology multiple-choice suite \citep{laurent2024labbench}. Every cell uses the same extraction pipeline and fixed \texttt{gpt-oss-120b} equivalence judge, so the Gemma actor never labels its own answers (Table~\ref{tab:cross_benchmark}; expanded interpretation in Appendix~\ref{app:cross_benchmark}). Three points stand out.

\begin{table}[t]
\centering
\small
\begin{tabular}{lrrrr}
\toprule
 & math & bio & bio & math \\
Actor & gpt-oss & gpt-oss & gemma & gemma \\
\midrule
Valid poll cand. & 23{,}391 & 2{,}912 & 4{,}852 & 14{,}129 \\
Candidate acc. & 0.518 & 0.593 & 0.491 & 0.491 \\
Mean stated conf. & 0.790 & 0.729 & 0.879 & 0.839 \\
ECE raw & 0.278 & 0.167 & 0.438 & 0.443 \\
ECE isotonic & 0.008 & 0.023 & 0.008 & 0.001 \\
AUROC raw & 0.721 & 0.637 & 0.537 & 0.440 \\
AUROC isotonic & 0.720 & 0.630 & 0.540 & 0.530 \\
Routing argmax & 0.497 & 0.532 & 0.399 & 0.362 \\
Routing random & 0.490 & 0.569 & 0.455 & 0.474 \\
Routing oracle & 0.549 & 0.622 & 0.609 & 0.566 \\
Net commit.\ shift & $-1.7$ & $+0.9$ & $+3.3$ & $+12.2$ \\
\bottomrule
\end{tabular}
\caption{Descriptive point estimates from the three-axis audit over a $2\times2$ actor$\times$benchmark grid. The first column is the primary result. Routing rows are per-block success over all-routeable blocks; net commitment shift is the unconditional poll-to-speak correctness shift in percentage points. Correctness labels in all cells use the same fixed \texttt{gpt-oss-120b} equivalence judge, so the Gemma columns are cross-family-labeled. Per-cell confidence intervals are not reproduced in this preprint.}
\label{tab:cross_benchmark}
\end{table}

The grid separates conclusions that transfer from effects that remain setting-dependent. First, overconfidence recurs in every cell, and separately cross-fitted isotonic procedures reduce ECE to $\le 0.023$. This is a scale correction, not evidence of a common ranking signal: raw AUROC is 0.721 and 0.637 in the two \texttt{gpt-oss} cells, but 0.537 and 0.440 in the Gemma cells. Candidate accuracy alone does not explain the weak Gemma confidence signal in these experiments, but these values do not establish equal general reasoning capability or a model-wide inability to introspect. Second, raw-confidence routing is not reliably beneficial. Argmax is close to random-valid selection on \texttt{gpt-oss}/math (0.497 versus 0.490), trails it on \texttt{gpt-oss}/biology (0.532 versus 0.569), and performs substantially worse in both Gemma cells (0.399 versus 0.455 and 0.362 versus 0.474). Third, commitment divergence is substantial in every cell (PSDR 0.11--0.27), but its unconditional correctness shift is negative only on \texttt{gpt-oss}/math ($-1.7$~pp) and positive in the other three cells ($+0.9$ to $+12.2$~pp). The decomposition generalizes; the direction and practical importance of each effect must be measured in the deployment setting.

\section{Discussion}

\paragraph{Where the audit localizes the main limitation.}
Confidence-routed broadcast has a higher same-task pass rate than the available \texttt{single\_agent} comparator in this corpus (0.892 versus 0.788) but uses much more compute (12.8$\times$ tokens, Appendix~\ref{app:single_agent_context}). Because this is not an equal-compute comparison, we treat it as context rather than a preference claim. In the primary cell, fixed-router counterfactuals show little selection-rule headroom, while the exploratory outcome model associates successful traces with frequent availability of correct polled candidates. In the Gemma cells, however, replacing raw-confidence argmax with random-valid selection materially improves routing. Candidate availability and selection quality can therefore both matter, depending on the actor and task.

\paragraph{Scope and common objections.}
Divergence can be a productive form of test-time compute \citep{snell2024scaling,shinn2023reflexion}. Here, speaking rather than retaining the poll answer is associated with an unconditional local shift of $-1.7$~pp on \texttt{gpt-oss}/math and positive shifts in the other three cells. Format pressure is one plausible explanation: the structured poll fixes an answer under a constrained contract, whereas the free-form speak step reopens generation. We treat that as a hypothesis. The local shift is neither an estimate nor an upper bound on the end-to-end effect of a commitment lock, because changing a public message would alter the subsequent transcript and later generations (Appendix~\ref{app:protocol_details}).

\paragraph{Design implications.}
Because the grid shows that the useful action can change across settings, design decisions should be keyed to measured quantities. (1) Measure both candidate availability and selection headroom. (2) Evaluate whether confidence improves within-block selection using PCCA or direct paired router counterfactuals; pooled AUROC alone is insufficient for routing. (3) Recalibrate before interpreting the score as a probability, while evaluating any calibrated policy end to end. (4) Test commitment scaffolding as an intervention rather than inferring its system-level effect from local poll-to-speak shifts. For evaluators, we suggest reporting routing, calibration, and commitment separately via the compact template in Appendix~\ref{app:trace_audit_card}.

\section{Limitations}\label{sec:limitations}

This is a trace audit of one broadcast protocol, anchored on \texttt{gpt-oss-120b} math traces and repeated on the $2\times2$ grid of \S\ref{sec:ablation}. We do not claim all multi-agent systems, topologies, or actors share this profile, and the direction of the commitment effect varies across cells. We do not run a commitment-lock ablation, calibrated-router rerun, or equal-compute single-agent baseline. The available same-task comparators (Appendix~\ref{app:single_agent_context}) use far less compute, so they provide context rather than a matched-cost frontier.

The equivalence judge is \texttt{gpt-oss-120b} in a self-judge configuration. Two Gemma 4 variants give the same sign for conditional NPCS when they replace the residual-case labels, but this does not identify which judge is correct or rule out errors shared by all judges. The derivation analysis also depends on the audited visibility model (peer polls excluded from speaker context), so the novel-vs-borrow split should not be read as model-general. Training-data overlap is not measured and can affect both pooled and tier-stratified estimates (Appendix~\ref{app:contamination_probe}).

\section{Conclusion}

A confidence-routed turn contains three measurement objects that should not be collapsed. In the primary \texttt{gpt-oss}/math cell, tested fixed routers fall within 1.1~pp, and a cross-fitted isotonic procedure reduces held-out ECE from $0.278$ to $0.008$. Across the grid, however, raw confidence can be weak or inverted: confidence argmax performs 5.6 and 11.2~pp below random-valid selection in the Gemma cells. Poll-to-speak divergence is likewise substantial across settings, but its unconditional correctness shift ranges from $-1.7$ to $+12.2$~pp. The transferable conclusion is to audit routing discrimination, probability calibration, and public commitment separately: scale repair cannot create a missing routing signal, and commitment effects must be measured rather than assumed.

\section*{Acknowledgments}

This research used resources of the Argonne Leadership Computing Facility, a
U.S. Department of Energy (DOE) Office of Science user facility at Argonne
National Laboratory (ANL) operated under Contract No.\ DE-AC02-06CH11357.

\small
\bibliography{references}
\normalsize

\appendix

\section*{Technical Appendix}

The appendix first supplies headline denominators and a worked trace, then documents protocol surfaces, estimands, robustness checks, judge sensitivity, cross-cell results, contamination limits, and the Trace-Audit Card. Section~\ref{app:analysis_artifacts} closes with the exact availability boundary for this preprint.

\section{Headline denominators and a worked trace}\label{app:body_floats}
This section gives the full headline-metrics table with explicit denominators (Table~\ref{tab:headline}) and a worked example that instantiates all three audit objects on one trace (Figure~\ref{fig:worked_example}).

\begin{table*}[t]
\centering
\small
\setlength{\tabcolsep}{4pt}
\begin{tabular}{llccc}
\toprule
Axis & Metric & Estimate & 95\% CI & Denominator \\
\midrule
Routing & CRA & 0.506 & [0.486, 0.527] & 9,467 sel-valid blocks \\
Routing & HRWP & 0.906 & [0.896, 0.915] & 5,295 possible-hit blocks \\
Routing & MCR & 0.058 & [0.052, 0.065] & 7,970 eligible blocks \\
Routing & PCCA (tie-aware) & 0.570 & [0.549, 0.593] & 1,184 blocks / 2,079 pairs \\
Routing & CTR & 0.327 & [0.305, 0.349] & 2,079 corr-vs-wrong pairs \\
Routing & Selected-invalid rate & 0.149 & [0.139, 0.159] & 11,122 selected blocks \\
Routing & Sel-invalid $\mid$ corr alt exists & 0.007 & [0.005, 0.011] & 5,295 possible-hit blocks \\
\midrule
Commitment & PSDR & 0.204 & [0.194, 0.214] & 9,436 valid pairs \\
Commitment & PSIR (cond.\ on divergence) & 0.174 & [0.153, 0.195] & 1,927 divergent pairs \\
Commitment & PSDRg (cond.\ on divergence) & 0.257 & [0.233, 0.283] & 1,927 divergent pairs \\
Commitment & $\mathrm{NPCS}_{\mid\mathcal D}$ & $-0.084$ & [$-0.117$, $-0.050$] & 1,927 divergent pairs \\
Commitment & Net shift (unconditional) & $-0.017$ & [$-0.024$, $-0.010$] & 9,436 valid pairs \\
\bottomrule
\end{tabular}
\caption{Headline metrics with explicit denominators and problem-cluster bootstrap 95\% CIs. Commitment metrics are reported both conditional on divergence and unconditional over all valid labeled selected pairs.}
\label{tab:headline}
\end{table*}

\begin{figure*}[!t]
\centering
\fbox{\begin{minipage}{0.94\textwidth}
\small
\textbf{Worked example: one trace, all three audit objects.}
\textit{Tier 03, source} \texttt{fermat}, \textit{problem id} \texttt{q0002}, \textit{gold answer} $\boxed{27}$.

\smallskip
\textbf{Problem context.} Competition-style number theory with reference answer $27$. The original wording and any close paraphrase are omitted because reuse terms are source-specific.

\smallskip
\textbf{Round 0, poll block 1.} Structured poll returned by each deliberator (one trace from the analysis record):
\begin{center}
\setlength{\tabcolsep}{4pt}
\begin{tabular}{lccc}
\toprule
Agent & \shortstack{polled\\cand.\ $c_i$} & \shortstack{conf.\ $s_i$\\(1--100)} & correct? \\
\midrule
A1 (Mira)  & $\mathbf{27}$ & 78 & \checkmark \\
A2 (Rowan) & $24$          & 78 &            \\
A3 (Talia) & $30$          & 78 &            \\
\bottomrule
\end{tabular}
\end{center}

\textbf{Routing.} All three confidences tie at $0.78$, and the runtime first-seen tie rule selects A1 (Mira). The pool contains exactly one correct candidate, so this is one of 5{,}295 HRWP-eligible blocks and one of the 90.6\% in which the router hits a correct candidate (Table~\ref{tab:headline}).

\textbf{Calibration.} The pool's mean stated confidence on this block is $0.78$ while its empirical accuracy is $1/3$, a $\sim$$45$~pp local overconfidence gap that contributes to the headline pooled $+27.2$~pp gap (main paper).

\textbf{Commitment.} A1 Mira's free-form public message at speak time ends with the boxed answer $24$, not the polled $27$. Because peer polls are not visible at speak time and no prior speak or system incumbent contained $24$, the audit classifies this divergence as \emph{novel} (see the derivation breakdown in the main paper); the answer that won the turn was correct, the answer that entered the transcript is wrong. This is a $1\!\to\!0$ shift of the kind aggregated by $\mathrm{NPCS}_{\mid\mathcal D}=-8.4$~pp.

\textbf{Trace outcome.} The trace later recovers: in round~1 the three agents converge on $27$ with confidences $78$--$85$, the speak step preserves the polled answer, and the broadcast finally returns $27$ (PASS). The single block above contributes one row to each of the three pooled metric families defined in the main paper: one HRWP-eligible block, one valid poll candidate per agent, and one divergent valid pair.
\end{minipage}}
\caption{One concrete trace through the three audit objects. The router selects A1 by first-seen tie-break on a triple-tied confidence; the polled answer that earned the turn ($27$) is not the answer A1 publicly states ($24$). The block, pair, candidate, and pool abstractions defined in the main paper all instantiate on this single trace, and the same trace later succeeds in round~1.}
\label{fig:worked_example}
\end{figure*}
\FloatBarrier

\section{Protocol and analysis details}\label{app:protocol_details}

\paragraph{Formal metric definitions (referenced in the main paper).}
For the headline pair-micro estimates, let $\mathcal{P}$ be the pooled set of valid labeled selected poll-speak pairs $(c^\poll, c^\speak)$ in an analysis cell, with each pair linked to problem $q$ and gold answer $y_q$. Let $\mathcal{D} = \{(c^\poll, c^\speak) \in \mathcal{P} : \answer(c^\poll) \not\equiv \answer(c^\speak)\}$ be the divergent subset. The commitment metrics are
\begin{align*}
\text{PSDR}    &= |\mathcal{D}| / |\mathcal{P}|, \\
\text{PSIR}    &= \tfrac{1}{|\mathcal{D}|}\big|\{(c^\poll, c^\speak) \in \mathcal{D} : \\
               &\qquad \neg\correct(c^\poll, q) \land \correct(c^\speak, q)\}\big|, \\
\text{PSDRg}   &= \tfrac{1}{|\mathcal{D}|}\big|\{(c^\poll, c^\speak) \in \mathcal{D} : \\
               &\qquad \correct(c^\poll, q) \land \neg\correct(c^\speak, q)\}\big|, \\
\text{NPCS}_{\mid\mathcal{D}} &= \text{PSIR} - \text{PSDRg}, \\
\text{NPCS}_\text{uncond} &= \tfrac{1}{|\mathcal{P}|} \sum_{(c^\poll, c^\speak) \in \mathcal{P}} \big[\correct(c^\speak,q) \\
 &\qquad - \correct(c^\poll,q)\big].
\end{align*}
The conditional $\mathrm{NPCS}_{\mid\mathcal D}$ is micro-averaged over divergent pairs, whereas $\mathrm{NPCS}_\text{uncond}$ is micro-averaged over all valid pairs. The exploratory outcome regression instead uses unconditional wrong-to-right and right-to-wrong event rates over selected speak events within each trace; zero events contribute zero. The block-level same-trace predictor used in that regression is
\begin{equation*}
\begin{aligned}
\text{pool-solvable}(q) = {} & \tfrac{1}{|\mathcal{B}_q|} \sum_{b \in \mathcal{B}_q} \ind\big\{\exists\, c \in b : \\
& \valid(c) \land \correct(c, q)\big\}
\end{aligned}
\end{equation*}
where $\mathcal{B}_q$ is the set of poll blocks for trace $q$ and $b$ enumerates valid candidates within a block. This depends on the agent pool's content, not on which candidate the router selected.

\paragraph{System incumbent (referenced in the main paper and in Table~\ref{tab:adoption_logit}).}
The \emph{system incumbent} is the protocol's currently endorsed candidate answer for a problem. It is set or updated by candidate-review / approval events emitted between speak rounds (logged as \texttt{approval\_*\_candidate\_update} in the trace), separately from any individual agent's poll or speak. Operationally, the audit records a per-trace \texttt{incumbent\_history} as a list of $((\text{round}, \text{block\_id}), \text{candidate})$ tuples, and the function \texttt{last\_incumbent\_at\_or\_before}($r, b$) returns the most recently endorsed candidate at or before key $(r, b)$. The incumbent therefore persists across rounds until the next approval event overrides it. The incumbent-pull class is the only divergence class with a non-negative point estimate ($+1.7$~pp), a pattern consistent with previously endorsed answers having accrued evidence; the observational taxonomy does not establish that endorsement caused the difference. Other broadcast variants without a system-tracked incumbent will not exhibit this class.

\begin{table*}[h]
\centering
\small
\setlength{\tabcolsep}{4pt}
\begin{tabular}{p{0.10\linewidth}p{0.29\linewidth}p{0.23\linewidth}p{0.29\linewidth}}
\toprule
Stage & Visible context & Output contract & Downstream role \\
\midrule
Poll & Problem, outer-loop evaluator feedback, current candidate answer, public discussion context, deferred private notes, own speaking history & Structured JSON with \texttt{score}, \texttt{reason}, \texttt{intent}, \texttt{candidate\_answer}, \texttt{has\_candidate\_answer} & Produces the candidate and confidence score used for turn allocation \\
Router & The per-agent poll outputs; runtime rule is confidence argmax with deterministic first-seen tie resolution & Selected speaker identity for the current turn & Allocates the floor; the top-confidence tie-random counterfactual reported in the main paper is the order-effect control \\
Speak & Selected agent sees the public transcript and system incumbent, but not peers' private poll JSON & Free-form public message, optionally with a boxed answer & Public commitment that enters the shared transcript and may diverge from the polled candidate \\
Evaluator & Final submitted answer plus evaluator-only reference material & PASS/FAIL, correctness bit, non-leaking repair hint & Supplies the external correctness signal used by the original run and some post-hoc labels \\
\bottomrule
\end{tabular}
\caption{Runtime protocol surface for the audited broadcast configuration. The central asymmetry is that the poll is structured and private to the router, while the speak step is free-form and public. This is the protocol fact behind the poll/speak commitment gap.}
\label{tab:protocol_surface}
\end{table*}

\begin{table}[h]
\centering
\small
\setlength{\tabcolsep}{4pt}
\begin{tabularx}{\linewidth}{p{0.32\linewidth}X}
\toprule
Setting & Audited value \\
\midrule
Deliberation pool & Three homogeneous \texttt{openai/gpt-oss-120b} peers at temperature 0, plus an evaluator from the same model family \\
Discussion / approval schedule & Four discussion turns per outer round, followed by two approval rounds \\
Routing rule & Confidence argmax with deterministic first-seen tie resolution \\
Deferred-note threshold & Non-selected agents write deferred private notes only if their poll score is at least 65 \\
Public-memory regime & \texttt{summary\_plus\_recent} with \texttt{recent\_history\_window = 5} \\
Approval / candidate update policy & Unanimous approval; reviewed candidate updated by \texttt{majority\_revision}; system incumbent persists until overridden \\
Reviewer feedback mode & Non-leaking evaluator \texttt{hint} fed into later outer rounds \\
Prompt budget caps per peer & Public context 2400 tokens, evaluator advice 2400, private notes 600, own speak history 600, candidate-answer field 300, total manual prompt 12000 \\
Evaluator output contract & \texttt{Verdict: PASS/FAIL}, \texttt{Correctness: 0/1}, plus a brief non-leaking repair hint \\
\bottomrule
\end{tabularx}
\caption{Load-bearing configuration knobs for the audited broadcast run. These settings matter because they determine what information the poll and speak stages actually see, how much context survives truncation, and how the system-level incumbent is updated between public speak events. They make the audit object more precise than ``prompting style'' alone.}
\label{tab:audited_config_knobs}
\end{table}

\paragraph{Difficulty-tier construction (referenced in the main paper).}
The 10 tiers come from the \texttt{difficulty\_tier} field of the source corpus (\path{omni-math-2-filtered/tier_NN.jsonl}), which assigns each problem to a tier based on its Omni-MATH difficulty score \citep{gao2024omnimath}. Tiers are ordinal: tier 01 is easiest, tier 10 is hardest, but the tier-to-tier difficulty step is not equal-interval. Pooled headline metrics use cluster-bootstrap over all 421 problem clusters and are not affected by tier scaling. Where regressions in main paper and Appendix~\ref{app:regression} use tier-difficulty as a continuous predictor, results should be read as monotone trends rather than as exact linear effects per tier-step; a sensitivity refit treating tier as a factor (rather than continuous) gives qualitatively identical sign and significance for the score $\times$ difficulty interaction in the calibration regression.

\paragraph{Statistical reporting parameters (referenced in main paper).}
Cluster-robust standard errors use \texttt{statsmodels.api.Logit} with \texttt{cov\_type=`cluster'} and \texttt{groups=problem\_cluster\_id}, corresponding to the CR1 estimator with the conventional small-sample $G/(G-1)$ correction. We did not refit with CR2/CR3, so small regression $p$-values should be read as Wald-asymptotic. Cluster-bootstrap CIs use the percentile method with $B=2000$ problem-cluster resamples and seed $20260502$; we did not implement BCa. Predictor lists were not pre-registered, and analyses are not adjusted for multiplicity. We therefore emphasize effect sizes and uncertainty intervals for headline descriptive comparisons and treat marginal regression $p$-values as hypothesis-generating.

\paragraph{Relation to confidence-visibility studies.}
\citet{wu2025reallydebate} study whether exposing agents to peers' verbalized confidence improves debate accuracy. Our protocol uses confidence differently: private poll scores are consumed by the router and are not shown to the speaker. The primary routing result therefore concerns within-block allocation, not peer conditioning. The adoption logit in Table~\ref{tab:adoption_logit} also is not a test of confidence visibility. Its source-confidence variable is analyst metadata attached to a previously visible candidate; it shows that candidates associated post hoc with higher source-poll confidence are more likely to match the spoken answer after controlling for the listed covariates. It does not establish that the speaker observed or responded to confidence, nor that confidence caused adoption.

\paragraph{What the local commitment shift does not identify.}
The audit measures revisions but does not run a commitment-lock ablation. Across valid selected events in \texttt{gpt-oss}/math, speaking rather than retaining the poll answer is associated with an unconditional local shift of $-1.7$~pp. This is not an estimate or upper bound on the end-to-end effect of a lock: replacing a public message would alter the subsequent transcript, later generations, and possibly the final answer. A system-level effect requires an intervention or rerun. Positive local shifts in the other three cells additionally caution against a blanket lock.

\section{Notation: metrics referenced throughout the paper}\label{app:notation}

Table~\ref{tab:notation} is a one-look reference to the abbreviations used in the body and the remaining appendices, with a pointer to the section that introduces each. Formal definitions, the validity-gating predicate $\valid(c)$, and the equivalence relation $\equiv$ are in Appendix~\ref{app:protocol_details}.

\begin{table}[h!]
\centering
\small
\setlength{\tabcolsep}{3pt}
\begin{tabular}{lp{0.55\linewidth}l}
\toprule
Symbol & Definition & Intro \\
\midrule
CRA   & Candidate Routing Accuracy: fraction of selected-valid blocks where the routed candidate is correct & main paper \\
HRWP  & Hit Rate When Possible: among blocks where any valid candidate is correct, fraction where the selected one is also correct & main paper \\
MCR   & Missed Correct Rate: fraction of blocks where the selected valid candidate is wrong but a non-selected valid candidate is correct & main paper \\
PCCA  & Pairwise Confidence-Correctness Alignment: block-macro average of each mixed block's tie-aware correct--wrong comparisons; $0.5$ means no within-block ordering signal & main paper \\
CTR   & Confidence Tie Rate: fraction of within-block (correct, wrong) candidate pairs that are tied in confidence & main paper \\
ECE   & Expected Calibration Error: weighted mean absolute gap between binned mean confidence and bin accuracy & main paper \\
MCE   & Maximum Calibration Error: maximum absolute gap across calibration bins & main paper \\
AUROC & Rank-discrimination of confidence: probability a correct poll candidate is scored above a wrong one, pooled over all valid candidates & main paper \\
PSDR  & Poll-Speak Divergence Rate: fraction of valid selected pairs where the polled and spoken answers disagree & main paper \\
PSIR  & Poll-Speak Improvement Rate, conditional on divergence: wrong-to-right transitions & main paper \\
PSDRg & Poll-Speak Degradation Rate, conditional on divergence: right-to-wrong transitions & main paper \\
$\mathrm{NPCS}_{\mid\mathcal D}$ & Conditional net shift among divergent pairs, $\mathrm{PSIR}-\mathrm{PSDRg}$ & main paper \\
$\mathrm{NPCS}_{\mathrm{unc}}$ & Unconditional net shift among all valid labeled pairs & main paper \\
\bottomrule
\end{tabular}
\caption{Abbreviation glossary for the body and appendix metrics. Routers and router-counterfactual names (\texttt{runtime\_argmax}, \texttt{top\_confidence\_tie\_random}, \texttt{random\_valid}, \texttt{majority\_vote}, \texttt{confidence\_weighted\_vote}, \texttt{oracle}) are defined in-context in the main paper; the pool-solvable rate used in the outcome-linkage regression is also defined there.}
\label{tab:notation}
\end{table}

\section{Per-tier descriptive metrics}\label{app:pertier}

Table~\ref{tab:pertier} reports routing, ranking, commitment, and calibration metrics by difficulty tier. Pooled values appear in Appendix Table~\ref{tab:headline}; this table is the per-tier complement for the trends underlying the calibration regression in Appendix~\ref{app:regression} and Figure~\ref{fig:tier_rank_vs_calibration}.

\begin{table}[h!]
\centering
\small
\setlength{\tabcolsep}{4pt}
\begin{tabular}{lrrrrrr}
\toprule
Tier & HRWP & PCCA & CTR & PSDR & $\mathrm{NPCS}_{\mid\mathcal D}$ & ECE \\
\midrule
01 & 0.991 & 0.500 & 0.000 & 0.008 & 0.000 & 0.103 \\
02 & 0.962 & 0.572 & 0.383 & 0.125 & 0.059 & 0.128 \\
03$^*$ & 0.853 & 0.575 & 0.368 & 0.170 & $-0.250$ & 0.181 \\
04 & 0.936 & 0.587 & 0.353 & 0.166 & $-0.086$ & 0.202 \\
05 & 0.915 & 0.602 & 0.361 & 0.201 & $+0.022$ & 0.292 \\
06 & 0.871 & 0.578 & 0.307 & 0.233 & $-0.093$ & 0.356 \\
07 & 0.868 & 0.544 & 0.321 & 0.226 & $-0.184$ & 0.369 \\
08 & 0.859 & 0.532 & 0.303 & 0.281 & $-0.133$ & 0.311 \\
09 & 0.843 & 0.522 & 0.227 & 0.239 & $-0.170$ & 0.409 \\
10$^*$ & 0.806 & 0.574 & 0.267 & 0.093 & $-0.400$ & 0.348 \\
\bottomrule
\end{tabular}
\caption{Per-tier routing, commitment, and calibration metrics. NPCS is conditional on divergent pairs. Tiers marked with $^*$ have small denominators and should be interpreted with caution: tier 03 has 2 problem clusters and 47 valid pairs; tier 10 has 2 clusters and 54 valid pairs. Headline pooled estimates use cluster bootstrap over all 421 problem clusters.}
\label{tab:pertier}
\end{table}

\begin{figure}[h]
\centering
\includegraphics[width=0.95\linewidth]{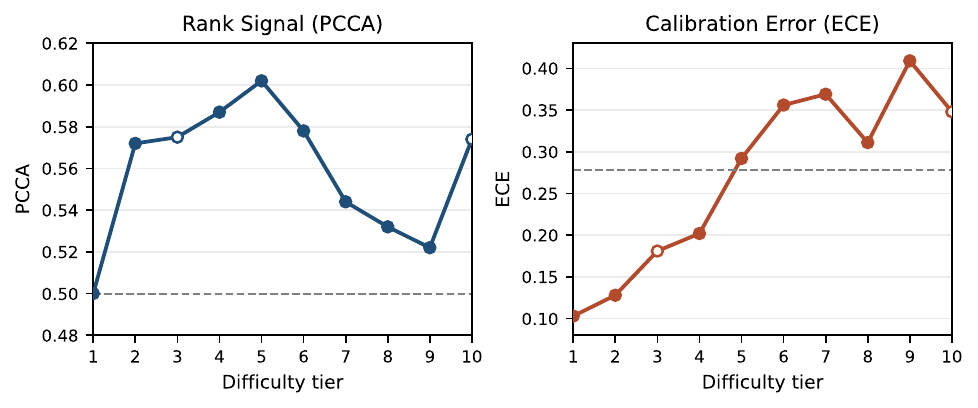}
\caption{Per-tier rank-versus-calibration split derived from Table~\ref{tab:pertier}. Left: PCCA stays above the 0.5 no-signal floor across most tiers, but weakens toward harder slices. Right: ECE climbs sharply with difficulty, reaching the 0.31--0.41 range on tiers 06--10. Hollow markers indicate the small-denominator tiers 03 and 10. The point of the figure is not that either curve is perfectly monotone; it is that the probability interpretation degrades faster than the rank signal.}
\label{fig:tier_rank_vs_calibration}
\end{figure}

\section{Regression highlights}\label{app:regression}

The pooled correctness and divergence logistic fits referenced from the main paper are reported in Table~\ref{tab:regression}. The per-event adoption logit underlying the novel-vs-borrow derivation breakdown (main paper) is reported in Table~\ref{tab:adoption_logit}, and the non-exclusive match-flag counts behind that breakdown in Table~\ref{tab:derivation_flags}. The outcome-linkage fit appears in the main paper; the pool-solvable / selected-correct + routing-slack decomposition of that fit is in Appendix~\ref{app:linkage_decomp}.

\begin{table*}[h!]
\centering
\small
\begin{tabular}{llrrrr}
\toprule
Model & Term & Coef. & 95\% CI & OR & $p$ \\
\midrule
Correctness & score (norm) & 6.551 & [5.844, 7.258] & 700.1 & $1.18\cdot 10^{-73}$ \\
Correctness & score $\times$ difficulty & $-1.993$ & [$-2.303$, $-1.683$] & 0.136 & $2.44\cdot 10^{-36}$ \\
Correctness & round index & $-0.903$ & [$-1.022$, $-0.785$] & 0.405 & $1.26\cdot 10^{-50}$ \\
\midrule
Divergence & poll correct & $-1.171$ & [$-1.333$, $-1.009$] & 0.310 & $1.23\cdot 10^{-45}$ \\
Divergence & poll correct $\times$ difficulty & 0.254 & [0.181, 0.328] & 1.290 & $8.85\cdot 10^{-12}$ \\
Divergence & score (norm) & $-1.208$ & [$-1.886$, $-0.530$] & 0.299 & $4.83\cdot 10^{-4}$ \\
Divergence & score $\times$ difficulty & 0.495 & [0.084, 0.906] & 1.640 & $0.018$ \\
\bottomrule
\end{tabular}
\caption{Cluster-robust pooled logistic regressions with agent fixed effects. Correctness model uses 23{,}391 valid poll candidates; divergence model uses 9{,}436 valid selected poll-speak pairs. Both use cluster-robust SEs by problem cluster ($n_{\text{clusters}} = 421$). The outcome-linkage regression reported in the main paper is fit on the 4{,}181-trace level.}
\label{tab:regression}
\end{table*}

\begin{table}[h!]
\centering
\small
\setlength{\tabcolsep}{2pt}
\begin{tabular}{lrrr}
\toprule
Term & Coef. & OR & $p$ \\
\midrule
role $=$ self-prior-speak (vs.\ peer)  & $-0.010$ & 0.99 & 0.89 \\
role $=$ incumbent (vs.\ peer)         & $+0.519$ & 1.68 & $8.1\cdot 10^{-3}$ \\
lag (rounds back)                      & $+0.004$ & 1.00 & 0.95 \\
source confidence (analyst covariate)  & $+1.471$ & 4.35 & $7.3\cdot 10^{-4}$ \\
source was correct (vs.\ gold)         & $+0.314$ & 1.37 & $4.2\cdot 10^{-4}$ \\
candidate is most-recent (1/0)         & $+0.198$ & 1.22 & $4.7\cdot 10^{-2}$ \\
candidate in visible majority (1/0) & $+0.811$ & 2.25 & $1.9\cdot 10^{-35}$ \\
problem difficulty (z)                 & $-0.011$ & 0.99 & 0.54 \\
\bottomrule
\end{tabular}
\caption{Adoption logit. One row per (divergent speak event, visible candidate), with \texttt{adopted=1} when the spoken answer matches that candidate. Cluster-robust SEs use \texttt{problem\_uid}. Source confidence is post-hoc metadata from the candidate's originating private poll; it was not visible to the speaker, so its coefficient is an association rather than a confidence-visibility effect.}
\label{tab:adoption_logit}
\end{table}

\begin{table}[h!]
\centering
\small
\setlength{\tabcolsep}{4pt}
\begin{tabular}{lrr}
\toprule
Match flag (non-exclusive) & N & \shortstack{\% of\\diverg.} \\
\midrule
\texttt{matches\_self\_revert}    & 367     & 19.0 \\
\texttt{matches\_incumbent\_pull} & 224     & 11.6 \\
\texttt{matches\_peer\_echo}      & 550     & 28.5 \\
No match (novel)                  & 1{,}202 & 62.4 \\
\bottomrule
\end{tabular}
\caption{Independent match flags over the same $N=1{,}927$ divergent valid pairs as the precedence-resolved primary-class table in the main paper. A single divergent spoken answer can fire more than one flag, so the counts sum to more than $1{,}927$ and the percentages to more than $100$. The flag-versus-primary gap sits almost entirely in peer-echo: a peer's prior public answer matches $28.5\%$ of divergences, but only $12.4\%$ survive as \emph{primary} peer-echo, because the other $312$ also match the speaker's own earlier answer or the system incumbent, which take precedence. Self-revert and novel are unaffected by precedence, so their flag and primary counts coincide. Read against the primary-class table, this bounds how much divergence could be attributed to peer anchoring under a maximally permissive attribution rule.}
\label{tab:derivation_flags}
\end{table}

\section{Robustness checks for the main claims}\label{app:robustness_overview}

This section indexes checks for the primary-cell router comparison, pooled calibration, conditional NPCS judge sensitivity, and the exploratory outcome model. Table~\ref{tab:robustness_overview} maps five reviewer concerns to the corresponding appendix evidence.

\begin{table*}[h!]
\centering
\small
\begin{tabular}{p{0.15\linewidth}p{0.27\linewidth}p{0.40\linewidth}p{0.08\linewidth}}
\toprule
Concern & Check & Main result & Where \\
\midrule
Confidence ranking is just noise; PCCA / CTR are artifacts of the metric & Within-block confidence permutation null and noisy-oracle synthetic on the same trace data, no model re-runs & Permutation drives PCCA to $0.500$ and changes CTR from $0.327$ to $0.300\pm0.006$; noisy oracle drives PCCA toward $1$ and CTR toward $0$. This validates PCCA's no-signal reference, not CTR invariance. & App.~Table~\ref{tab:sanity} \\
\addlinespace
Self-judge artifact: \texttt{gpt-oss-120b} judges its own outputs and may bias NPCS & Two-judge relabeling of 4{,}379 residual pairs with \texttt{google/gemma-4-31B-it} ($\kappa{=}0.82$) and \texttt{google/gemma-4-E4B-it} ($\kappa{=}0.75$) & Conditional NPCS over the 9{,}263-pair tiers 02--09 population has the same negative sign under all three residual-label variants ($-0.083$, $-0.061$, $-0.060$). & App.~\ref{app:judge_audit} \\
\addlinespace
Outcome linkage depends on a single regression specification & Fit a related linkage logit with selected-correct rate and routing slack entered separately, using the same remaining predictors and cluster-robust SEs & The same-trace content variables remain strongly associated with outcome, but neither specification is interpreted causally. & App.~\ref{app:linkage_decomp} \\
\addlinespace
Single benchmark only (olympiad math throughout) & Repeat the audit on LAB-Bench biology \citep{laurent2024labbench} for both actors, using the same extraction pipeline and fixed judge & Overconfidence and scale repair recur, while routing and commitment effects differ: raw-confidence argmax trails random-valid selection, and unconditional commitment shifts are positive in the biology cells. & App.~\ref{app:cross_benchmark} \\
\addlinespace
Single actor family only (\texttt{gpt-oss-120b} deliberators throughout) & Cross-family replication with \texttt{google/gemma-4-31B-it} deliberators on both benchmarks and a fixed \texttt{gpt-oss-120b} judge & Gemma polls parse at 98.4\%; scale correction lowers ECE, while raw AUROC remains weak (0.537) or inverted (0.440), and raw-confidence argmax trails random-valid selection. & App.~\ref{app:cross_benchmark} \\
\bottomrule
\end{tabular}
\caption{High-priority robustness and scope checks. The table distinguishes descriptive sensitivity from causal or deployment claims; no row turns a same-trace association into an intervention estimate.}
\label{tab:robustness_overview}
\end{table*}

\section{Sanity-check constructions}\label{app:sanity}

We recompute the routing metrics under two synthetic constructions over the same trace data, without rerunning the deliberation traces. In the within-block permutation null, confidence scores are shuffled within each block while correctness labels stay fixed. In the noisy-oracle construction, each candidate's confidence becomes clipped\,$(\correct(c,q) + \mathcal{N}(0,\sigma))$ for $\sigma \in \{0.05, 0.10, 0.20, 0.50\}$. The null drives PCCA to its no-signal floor, while the noisy oracle drives PCCA toward 1 and CTR toward 0. CRA remains capped by the underlying content ceiling.

\begin{table*}[h]
\centering
\small
\setlength{\tabcolsep}{4pt}
\begin{tabular}{lrrrr}
\toprule
Construction & CRA & HRWP & PCCA & CTR \\
\midrule
Runtime (headline reproduction) & 0.506 & 0.906 & 0.570 & 0.327 \\
A: permutation null ($n{=}50$ reps) & 0.520 \footnotesize$\pm$.002 & 0.762 \footnotesize$\pm$.003 & \textbf{0.500} \footnotesize$\pm$.011 & 0.300 \footnotesize$\pm$.006 \\
B: noisy oracle, $\sigma{=}0.05$ & 0.622 \footnotesize$\pm$.001 & \textbf{1.000} & \textbf{1.000} & 0.000 \\
B: noisy oracle, $\sigma{=}0.10$ & 0.620 \footnotesize$\pm$.001 & \textbf{1.000} & \textbf{1.000} & 0.000 \\
B: noisy oracle, $\sigma{=}0.20$ & 0.621 \footnotesize$\pm$.001 & \textbf{1.000} & \textbf{1.000} & 0.000 \\
B: noisy oracle, $\sigma{=}0.50$ & 0.605 \footnotesize$\pm$.001 & 0.961 \footnotesize$\pm$.002 & 0.913 \footnotesize$\pm$.009 & 0.023 \footnotesize$\pm$.004 \\
\bottomrule
\end{tabular}
\caption{Sanity-check constructions over the existing traces. The permutation null confirms that runtime PCCA is meaningfully above the no-signal floor, while the noisy oracle confirms that PCCA saturates under a clean signal. CRA remains capped because selected-valid accuracy cannot exceed the underlying content ceiling.}
\label{tab:sanity}
\end{table*}

\section{LLM-judge sensitivity audit (two-judge cross-check)}
\label{app:judge_audit}

The main-text metrics depend on a single LLM equivalence judge (\texttt{openai/gpt-oss-120b} in a self-judge configuration). To audit this dependency we re-judge every non-trivial residual pair (boxed strings differ; $n=4{,}379$ across tiers 02--09) with two Google Gemma 4 variants: \texttt{google/gemma-4-31B-it} and \texttt{google/gemma-4-E4B-it}. Public model cards are \url{https://huggingface.co/openai/gpt-oss-120b}, \url{https://huggingface.co/google/gemma-4-31B-it}, and \url{https://huggingface.co/google/gemma-4-E4B-it}. The Gemma judges share neither training organization nor model family with the original judge, but they are not independent of each other.

Both Gemma judges return labels on $\geq 90\%$ of residual pairs. The remainder are JSON-parse failures and are treated as ``divergent'' in the conservative recomputation below. Agreement with \texttt{gpt-oss-120b} is substantial: Cohen's $\kappa=0.816$ for the 31B judge and $0.748$ for the E4B judge (Table~\ref{tab:judge_agreement}). On the stratified 200-pair sample where all three judges decide, the two Gemma variants agree with each other at 94.4\%, compared with 92.7\% and 88.2\% agreement with \texttt{gpt-oss}. This shared asymmetry is a sensitivity pattern, not ground truth: without human adjudication it cannot identify which judge is correct or separate a \texttt{gpt-oss} tendency from correlated Gemma-family bias.

\begin{table*}[h]
\centering
\small
\begin{tabular}{lcccc}
\toprule
Judge pair & Pairs decided & Agreement & Cohen's $\kappa$ & Confusion (judge1: gpt-oss / judge2) \\
\midrule
gpt-oss vs gemma-4-31B-it & 4{,}048 & 0.908 & 0.816 & 1776/311/62/1899 \\
gpt-oss vs gemma-4-E4B-it & 3{,}945 & 0.873 & 0.748 & 1670/433/68/1774 \\
\bottomrule
\end{tabular}
\caption{Two-judge cross-check on full residual corpus. Confusion entries are (judge1=div, judge2=div) / (judge1=div, judge2=equ) / (judge1=equ, judge2=div) / (judge1=equ, judge2=equ). Both Gemma judges show an asymmetric disagreement pattern: when gpt-oss flagged a pair as divergent, the Gemma judge ``rescues'' it as equivalent more often (311 / 433) than the opposite direction (62 / 68).}
\label{tab:judge_agreement}
\end{table*}

\begin{figure}[h]
\centering
\includegraphics[width=0.82\linewidth]{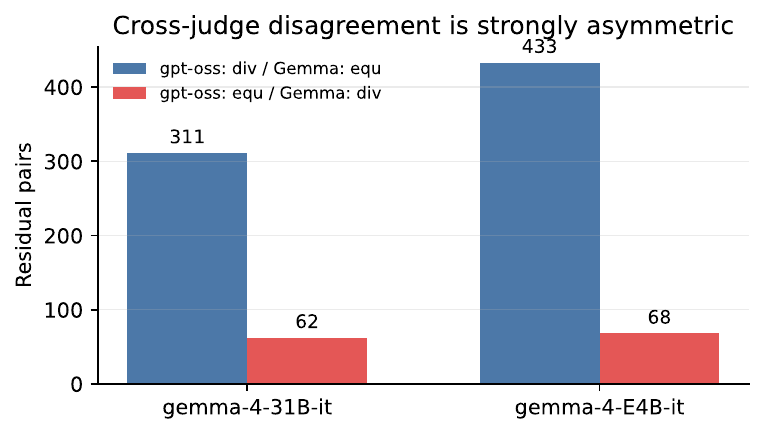}
\caption{Asymmetric disagreement in the cross-judge audit. On residual pairs where \texttt{gpt-oss} and Gemma disagree, the dominant observed direction is \emph{gpt-oss says divergent, Gemma says equivalent}. The figure describes disagreement direction; it does not determine which label is correct.}
\label{fig:judge_asymmetry}
\end{figure}

Table~\ref{tab:judge_sensitivity} recomputes commitment metrics over tiers 02--09, whose base population is 9{,}263 valid pairs rather than the 9{,}436-pair headline population; tiers 01 and 10 contribute the omitted 173 pairs. Parse failures from a Gemma judge are conservatively treated as divergent, which can increase its PSDR. Each conditional rate uses that judge variant's divergent subset: 1{,}921 pairs for \texttt{gpt-oss}, 2{,}418 for Gemma 31B, and 2{,}537 for Gemma E4B. Conditional NPCS retains a negative sign across variants ($-0.083$, $-0.061$, and $-0.060$), while its magnitude and the divergent denominator vary.

\begin{table}[h]
\centering
\small
\setlength{\tabcolsep}{3pt}
\begin{tabular}{lccc}
\toprule
Metric & \shortstack{gpt-oss-120b\\tiers 02--09} & \shortstack{gemma-4\\31B-it} & \shortstack{gemma-4\\E4B-it} \\
\midrule
Divergent pairs & 1{,}921 & 2{,}418 & 2{,}537 \\
PSDR & 0.207 & 0.261 & 0.274 \\
PSIR (cond.\ divergence) & 0.174 & 0.132 & 0.126 \\
PSDRg (cond.\ divergence) & 0.257 & 0.192 & 0.186 \\
$\mathrm{NPCS}_{\mid\mathcal D}$ & $-0.083$ & $-0.061$ & $-0.060$ \\
\bottomrule
\end{tabular}
\caption{Commitment metrics under three residual-case label variants over the same 9{,}263 valid pairs from tiers 02--09. PSIR, PSDRg, and $\mathrm{NPCS}_{\mid\mathcal D}$ use the judge-specific divergent denominator shown in the first row. Conditional NPCS has the same negative sign across variants; this is a sensitivity check, not ground-truth adjudication.}
\label{tab:judge_sensitivity}
\end{table}

\section{Outcome-linkage decomposition: pool-solvable rate split into router-picked and routing-slack pieces}\label{app:linkage_decomp}

The headline outcome-linkage regression uses a block-level same-trace predictor: \emph{pool-solvable rate} = per-trace mean of $\ind\{\text{any valid candidate is correct in the block}\}$. A related specification enters two separately z-scored variables: the piece the runtime router picked (\emph{selected-correct rate}) and the remaining \emph{routing slack} (pool-solvable rate minus selected-correct rate). This is a diagnostic decomposition, not an algebraically equivalent regression after standardization.

\begin{table}[h]
\centering
\small
\setlength{\tabcolsep}{3pt}
\begin{tabular}{lrrr}
\toprule
Predictor (z-scored) & $\beta$ & \shortstack{OR\\per SD} & $p$ \\
\midrule
Selected-correct rate & $+5.94$ & $\approx 380$ & $3.1 \cdot 10^{-11}$ \\
Uncond. improvement-event rate & $+1.40$ & $4.08$ & $0.002$ \\
Routing slack (oracle $-$ argmax) & $+1.08$ & $2.96$ & $9.0 \cdot 10^{-4}$ \\
Tier difficulty & $+0.42$ & $1.52$ & $2.9 \cdot 10^{-4}$ \\
Selected-invalid rate & $+0.60$ & $1.81$ & $9.1 \cdot 10^{-6}$ \\
Divergence rate (PSDR) & $+0.19$ & $1.21$ & $0.10$ \\
Uncond. degradation-event rate & $-0.26$ & $0.77$ & $0.74$ \\
Mean selected score & $-0.14$ & $0.87$ & $0.46$ \\
Speak events ($\log 1+\cdot$) & $-1.62$ & $0.20$ & $1.6 \cdot 10^{-21}$ \\
\bottomrule
\end{tabular}
\caption{Related decomposition specification with selected-correct rate and routing slack entered separately. Selected-correct rate is mechanically closer to the broadcast outcome than pool-solvable rate. McFadden pseudo-$R^2$ is 0.802 here and 0.798 in the collapsed specification. Both use same-trace summaries and are interpreted descriptively.}
\label{tab:linkage_decomp}
\end{table}

\paragraph{Robustness to near-separation.}
Pool-solvable rate nearly separates the outcome in the reported model (OR $\approx121$, wide CI), so plain-MLE Wald standard errors can be distorted. We refit the collapsed specification using a Firth penalty and, separately, after dropping pool-solvable rate. The unconditional improvement-event rate remains positive in both (OR $3.5$ and $9.1$), and the Firth pool-solvable point estimate is similar where included (OR $\approx111$). Mean selected confidence spans OR 0.84--1.11 across specifications, so even its sign is not stable; the reported cluster-robust MLE is small and non-significant (OR 0.84, $p=0.35$). We do not interpret it as a robust marginal predictor after the listed same-trace covariates are included.

\section{Cross-actor and cross-benchmark replication}\label{app:cross_benchmark}

This section expands the cross-actor and cross-benchmark ablation in the main paper. The primary cell uses \texttt{openai/gpt-oss-120b} deliberators on the Omni-MATH-style math corpus. The $2\times2$ grid crosses that actor and \texttt{google/gemma-4-31B-it} with the math corpus and LAB-Bench biology multiple-choice suite \citep{laurent2024labbench}. Every cell uses the same poll/speak extraction pipeline, fixed \texttt{gpt-oss-120b} equivalence judge, cell-specific $K=2$ cross-fitted calibrators, and the same router counterfactual definitions. Holding the judge fixed while varying the actor avoids actor self-labeling in the Gemma cells, but does not make the fixed judge ground truth. Table~\ref{tab:cross_benchmark_supp} reports descriptive point estimates; per-cell confidence intervals and underlying analysis files are not included in this preprint.

\begin{table}[h]
\centering
\small
\begin{tabular}{lrrrr}
\toprule
 & math & bio & bio & math \\
Actor & gpt-oss & gpt-oss & gemma & gemma \\
\midrule
Valid poll cand. & 23{,}391 & 2{,}912 & 4{,}852 & 14{,}129 \\
Candidate acc. & 0.518 & 0.593 & 0.491 & 0.491 \\
Mean stated conf. & 0.790 & 0.729 & 0.879 & 0.839 \\
\addlinespace
ECE raw & 0.278 & 0.167 & 0.438 & 0.443 \\
ECE isotonic & 0.008 & 0.023 & 0.008 & 0.001 \\
AUROC raw & 0.721 & 0.637 & 0.537 & 0.440 \\
AUROC isotonic & 0.720 & 0.630 & 0.540 & 0.530 \\
\addlinespace
Routing argmax & 0.497 & 0.532 & 0.399 & 0.362 \\
Routing random & 0.490 & 0.569 & 0.455 & 0.474 \\
Routing oracle & 0.549 & 0.622 & 0.609 & 0.566 \\
\addlinespace
Net commit.\ shift & $-1.7$ & $+0.9$ & $+3.3$ & $+12.2$ \\
\bottomrule
\end{tabular}
\caption{Descriptive point estimates over the $2\times2$ actor$\times$benchmark grid. The first column is the primary result. Routing rows are per-block success over all-routeable blocks; net commitment shift is the unconditional poll-to-speak correctness shift in percentage points. Correctness labels in all cells use the same fixed \texttt{gpt-oss-120b} equivalence judge. Per-cell confidence intervals are not reproduced in this preprint.}
\label{tab:cross_benchmark_supp}
\end{table}

\paragraph{Calibration: scale repair recurs, but discrimination does not.} Across the four cells, raw confidence is overconfident (ECE $0.167$--$0.443$, with stated confidence exceeding accuracy by $13.6$--$38.8$~pp), and the same cross-fitted isotonic procedure, fit separately within each actor--benchmark cell, reduces ECE to $\le0.023$ on held-out candidates. This is a scale correction, not evidence of a common ranking signal. In the \texttt{gpt-oss} cells, pooled AUROC changes from 0.721 to 0.720 on math and from 0.637 to 0.630 on biology. In the Gemma cells, raw AUROC is 0.537 on biology and 0.440 on math; calibrated AUROC is 0.540 and 0.530. The calibrated Gemma scores also approach the observed base rate (mean calibrated confidence $\approx0.49$, accuracy $\approx0.49$, and Brier $\approx0.246$). Each fitted isotonic map is nondecreasing, so it cannot reverse strict order among values transformed by that map, but it may introduce ties; cross-fitting and cell-specific fitting also preclude treating the result as one shared map. The supported cross-cell conclusion is that post-hoc calibration improves the numerical probability scale on these held-out data, not that it supplies missing routing discrimination.

\paragraph{Routing: raw-confidence argmax is not reliably beneficial.} The grid exhibits two routing regimes. On \texttt{gpt-oss}/math, confidence argmax and random-valid selection are nearly tied (0.497 versus 0.490), with an oracle ceiling of 0.549. On \texttt{gpt-oss}/biology, argmax trails random-valid selection (0.532 versus 0.569), while the oracle reaches 0.622. The gap is larger in the Gemma cells: argmax scores 0.399 versus 0.455 for random-valid selection on biology and 0.362 versus 0.474 on math, with oracle ceilings of 0.609 and 0.566. Ignoring the raw scalar therefore recovers 5.6 and 11.2~pp in the two Gemma cells. The remaining oracle gaps show that candidate availability and selection quality can both constrain performance. These point estimates support verifying within-block discrimination before using raw confidence for routing; the preprint does not report per-cell intervals for significance claims.

\paragraph{Commitment: divergence recurs, but its direction varies.} The primary \texttt{gpt-oss}/math cell has an unconditional net shift of $-1.7$~pp; the other three cells have positive point estimates from $+0.9$ to $+12.2$~pp, with the largest on Gemma/math. Thus the poll is not always the public commitment (PSDR 0.11--0.27), but revision is not uniformly harmful. The decomposition generalizes more reliably than the sign of the local shift.

\section{Task-level single-agent context}\label{app:single_agent_context}

The main paper audits only broadcast traces, so all routing, calibration, and commitment metrics are defined on that protocol alone. A natural reviewer question is whether the audited broadcast system buys anything over a non-collaborative alternative on the same task family. The broader verifier-gated math evaluation used to generate these broadcast traces also includes \texttt{baseline\_llm}, \texttt{single\_agent}, and \texttt{broadcast} runs over the same 4{,}181 problems. Table~\ref{tab:single_agent_context} reports weighted task-level aggregates. These are useful comparators, but they are not equal-compute baselines for the present audit.

\begin{table}[!htbp]
\centering
\small
\setlength{\tabcolsep}{4pt}
\begin{tabular}{lcccc}
\toprule
Mode & \shortstack{Pass\\rate} & \shortstack{Avg.\\tokens} & \shortstack{Avg.\\calls} & \shortstack{Avg.\\wall (s)} \\
\midrule
\texttt{baseline\_llm} & 0.568 & 18.4K & 9.69 & 43.6 \\
\texttt{single\_agent} & 0.788 & 48.1K & 18.0 & 74.3 \\
\texttt{broadcast} & 0.892 & 616.5K & 134.6 & 294.9 \\
\bottomrule
\end{tabular}
\caption{Task-level context from the same broader verifier-gated math evaluation over the same 4{,}181 problems. Here \texttt{baseline\_llm} is a single direct LLM solve, \texttt{single\_agent} is one LLM with multi-turn self-revision, and \texttt{broadcast} is the multi-agent protocol audited in the body. Broadcast outperforms the same-task \texttt{single\_agent} comparator in final pass rate (0.892 vs.\ 0.788), but at much higher average cost: 12.8$\times$ tokens, 7.5$\times$ model calls, and 4.0$\times$ wall time. These rows are contextual protocol comparisons, not equal-compute frontiers for the broadcast trace audit. Aggregates are weighted across tier-level summaries from that evaluation.}
\label{tab:single_agent_context}
\end{table}
\FloatBarrier

\section{Future work: source-stratified contamination probe}\label{app:contamination_probe}

The pooled ECE and AUROC describe this corpus without relying on a causal interpretation of the tier trend. Training-data overlap may nevertheless affect both pooled and tier-stratified estimates, and it is not measured in this study. The descriptive increase in ECE across difficulty tiers can reflect intrinsic difficulty, uneven pre-training coverage, or both. Source and year metadata alone would be an imperfect contamination proxy because competition problems may be indexed online independently of edition year. A stronger external-validity test would use a held-out post-cutoff set designed to minimize public-index overlap and rerun the audit pipeline; that experiment is outside the present study.

\section{Analysis and material availability}\label{app:analysis_artifacts}

This arXiv source package contains only the manuscript source, bibliography, style files, and rendered figure assets needed to compile the paper. It does not include trace data, benchmark text, runtime code, prompt templates, per-cell bootstrap files, or an external artifact locator. Internal file and directory names are used only to describe provenance and should not be read as claims that an artifact is publicly available.

\paragraph{Licensing and source note.}
The underlying math problems come from competition-style sources with heterogeneous reuse terms, including HMMT-, Putnam-, USAMO-, APMO-, and Omni-MATH-linked metadata. No competition problem corpus or trace dataset is redistributed in this arXiv source package. Any separate distribution of benchmark text or derived traces requires its own source-license, participant, coauthor, and institutional review; source attribution fields alone do not grant redistribution rights.

\paragraph{Compute note.}
Appendix~\ref{app:single_agent_context} reports aggregate tokens, model calls, and wall time for \texttt{baseline\_llm}, \texttt{single\_agent}, and \texttt{broadcast} on the same 4{,}181 problems. The underlying per-example accounting records are not part of this preprint package.

\section{Trace-Audit Card v0.1}\label{app:trace_audit_card}

The trace-audit card is a structured report of a confidence-routed deliberation protocol's measurement properties, intended as a protocol-level analogue of model cards \citep{mitchell2019modelcards} and data cards \citep{pushkarna2022datacards}. Required fields (R) are minimum reporting. When exact prompts, counts, or uncertainty estimates cannot be distributed, the card must state that limitation rather than imply availability.

\paragraph{How to populate the card.}
Adapt \texttt{routing\_rule} and \texttt{visibility\_at\_speak} for peer-to-peer or learned routing, and state explicitly if speakers see peer polls. For heterogeneous pools, list every \texttt{agent\_model} and report calibration per model. Give a denominator for every metric, and label unavailable prompts, source counts, or uncertainty estimates rather than silently omitting them.

\begin{table*}[!t]
\centering
\small
\setlength{\tabcolsep}{4pt}
\begin{tabular}{p{0.29\linewidth}p{0.06\linewidth}p{0.58\linewidth}}
\toprule
Field & R/O & Description \\
\midrule
\multicolumn{3}{l}{\emph{Protocol identity}} \\
\texttt{protocol\_name} & R & Short name for the deliberation protocol. \\
\texttt{protocol\_version} & R & Version tag. \\
\texttt{paper\_of\_record} & O & Citation for protocol definition. \\
\midrule
\multicolumn{3}{l}{\emph{Inference setup}} \\
\texttt{agent\_model} & R & Model id(s); homogeneous or heterogeneous. \\
\texttt{pool\_size} & R & Number of agents per round. \\
\texttt{rounds\_per\_problem} & R & Maximum / typical. \\
\texttt{topology} & R & Broadcast / peer-to-peer / star / other. \\
\midrule
\multicolumn{3}{l}{\emph{Prompts and signals}} \\
\texttt{poll\_prompt} & R & Verbatim template, source-package path, hash + public locator, or explicit unavailable status. \\
\texttt{speak\_prompt} & R & Same disclosure rule for the public speak step. \\
\texttt{confidence\_field} & R & Name and range of the confidence signal in the poll output. \\
\texttt{routing\_rule} & R & e.g.\ \texttt{confidence\_argmax}, \texttt{majority\_vote}, \texttt{learned}. \\
\texttt{visibility\_at\_speak} & R & What the speaker sees when generating the public message (prior speaks only / prior speaks + peer polls / etc.). \\
\midrule
\multicolumn{3}{l}{\emph{Validity and equivalence}} \\
\texttt{validity\_rule} & R & How invalid candidates are detected (parse failure, no answer field, etc.). \\
\texttt{equivalence\_checker} & R & Numeric / symbolic fast paths and LLM judge identity. \\
\texttt{judge\_cross\_check} & O & Vendor(s) used for sensitivity, with per-judge $\kappa$. \\
\midrule
\multicolumn{3}{l}{\emph{Headline metrics (denominators required; uncertainty stated where available)}} \\
\texttt{routing.cra}, \texttt{.hrwp}, \texttt{.mcr}, \texttt{.pcca}, \texttt{.ctr}, \texttt{.selected\_invalid\_rate} & R & Per main paper. \\
\texttt{calibration.ece}, \texttt{.brier}, \texttt{.auroc}, \texttt{.mean\_conf}, \texttt{.accuracy} & R & ECE bin scheme + $K$ disclosed. \\
\texttt{commitment.psdr}, \texttt{.psir}, \texttt{.psdrg}, \texttt{.npcs\_cond}, \texttt{.npcs\_unc} & R & Conditional and unconditional estimands labeled separately. \\
\texttt{outcome\_linkage.\allowbreak pool\_solvable\_or} & O & Exploratory same-trace association, labeled non-causal. \\
\midrule
\multicolumn{3}{l}{\emph{Cohort and contamination}} \\
\texttt{corpus} & R & Source distribution and counts, or a summary plus explicit availability limitation. \\
\texttt{train\_data\_overlap} & O & Stratified ECE / accuracy by likely train-data presence. \\
\bottomrule
\end{tabular}
\caption{Trace-Audit Card v0.1 schema. The card describes the measurement properties of a deliberation protocol; it does not describe a single model (model cards do that) or a dataset (data cards do that).}
\label{tab:trace_audit_card_schema}
\end{table*}
\FloatBarrier

\begin{figure*}[!t]
\centering
\begin{minipage}{0.92\linewidth}
{\footnotesize
\begin{verbatim}
protocol_name: broadcast_confidence_routed_v1
protocol_version: 1.0
paper_of_record: this paper
agent_model: openai/gpt-oss-120b (homogeneous pool)
pool_size: 3 homogeneous peers per round, temperature 0
rounds_per_problem: max 3 outer rounds; 4 discussion + 2 approval turns/round
topology: broadcast (all agents see public transcript)
poll_prompt: structured JSON contract summarized here; verbatim template not
             distributed in this arXiv source package
speak_prompt: free-form public-message contract summarized here; verbatim
              template not distributed in this arXiv source package
confidence_field: poll.score in [1,100], normalized to [0,1]
routing_rule: confidence_argmax with deterministic first-seen tie rule
visibility_at_speak: prior-round public speak messages + system incumbent;
                     peer polls NOT visible
validity_rule: extractable boxed/numeric answer present
equivalence_checker: numeric + symbolic fast paths, then openai/gpt-oss-120b
judge_cross_check: google/gemma-4-31B-it (kappa=0.82; n=4048 decided pairs),
                   google/gemma-4-E4B-it (kappa=0.75; n=3945 decided pairs)
routing.cra: 0.506 [0.486, 0.527] (n=9467)
routing.hrwp: 0.906 [0.896, 0.915] (n=5295)
routing.mcr: 0.058 [0.052, 0.065] (n=7970 eligible blocks)
routing.pcca: 0.570 [0.549, 0.593] (block-macro; 1184 blocks,
              2079 correct-wrong comparisons)
routing.ctr: 0.327 [0.305, 0.349] (n=2079 correct-wrong comparisons)
routing.selected_invalid_rate: 0.149 [0.139, 0.159] (n=11122)
calibration.ece: 0.278 [0.270, 0.286], 10 equal-width bins (n=23391 candidates)
calibration.brier: 0.302 [0.290, 0.313] (n=23391 candidates)
calibration.auroc: 0.721 [0.708, 0.733] (n=23391 candidates)
calibration.mean_conf: 0.790; calibration.accuracy: 0.518 (n=23391)
commitment.psdr: 0.204 [0.194, 0.214] (n=9436 valid pairs); judge-range 0.20-0.27
commitment.psir: 0.174 [0.153, 0.195] (n=1927 divergent pairs)
commitment.psdrg: 0.257 [0.233, 0.283] (n=1927 divergent pairs)
commitment.npcs_cond: -0.084 [-0.117, -0.050] (n=1927)
commitment.npcs_unc: -0.017 [-0.024, -0.010] (n=9436)
outcome_linkage.pool_solvable_or: 121 [32,456], n=4181; exploratory,
                                  same-trace, non-causal
corpus: 4181 traces, 421 clusters, 64 sources; largest HMMT_2=1321 and
        HMMT_11=860; complete per-source counts not in this preprint
train_data_overlap: not evaluated in this study
\end{verbatim}
}
\end{minipage}
\caption{Filled Trace-Audit Card for the primary \texttt{gpt-oss}/math protocol. The card fills the schema's required fields and explicitly marks prompt, source-count, and overlap information not distributed in the arXiv package. All required headline metrics give explicit denominators. Unless otherwise labeled, bracketed metric intervals are problem-cluster bootstrap 95\% intervals; the outcome-linkage odds ratio uses a cluster-robust Wald 95\% interval.}
\label{fig:filled_card}
\end{figure*}
\FloatBarrier

\end{document}